\documentclass{article}

\usepackage{arxiv}

\usepackage[numbers]{natbib}

\usepackage{algorithm}
\usepackage{algorithmic}
\usepackage{hyperref}
\usepackage{url}
\usepackage{mathtools}
\usepackage{booktabs} 
\usepackage{makecell}
\usepackage{xcolor}
\usepackage[capitalize,noabbrev]{cleveref}

\usepackage{amsmath,amssymb,amsfonts,amsbsy,amsthm}
\usepackage{graphicx}
\usepackage{textcomp}
\usepackage{xcolor}
\usepackage{bm}
\usepackage{caption}
\usepackage{subcaption}
\usepackage{multirow}
\usepackage{latexsym}
\usepackage{ifthen}
\usepackage{tabularx}
\usepackage[normalem]{ulem}
\usepackage{xspace}
\usepackage{paralist}	
\usepackage{breqn}
\usepackage{color}
\usepackage[utf8]{inputenc}
\usepackage{arydshln}

\newtheorem{definition}{Definition}
\newtheorem{remark}{Remark}
\newtheorem{assumption}{Assumption}

\newcommand{\bx}{{\bm x}}

\newcommand{\bv}{{\bm v}}
\newcommand{\bz}{{\bm z}}
\newcommand{\bw}{{\bm w}}
\newcommand{\bu}{{\bm u}}
\newcommand{\by}{{\bm y}}  
  
\newcommand{\ad}{{\mathcal{A}}}   
\newcommand{\xf}{{\widehat{\bm x}^{+}}}
\newcommand{\xp}{{\widehat{\bm x}^{-}}}
\newcommand{\pf}{{\bm P}^{+}}
\newcommand{\pp}{{\bm P}^{-}}
\newcommand{\bk}{\mathcal{K}}  
\newcommand{\on}{\Omega}  
\newcommand{\cn}{\overline{\Omega}}  
\newcommand{\ef}{{\bm \epsilon }^{+}}
\newcommand{\ep}{{\bm \epsilon }^{-}}

\title{On observability and optimal gain design for distributed linear filtering and prediction}

\author{ Subhro Das \\
MIT-IBM Watson AI Lab, IBM Research\\
Cambridge, MA, USA \\
\texttt{subhro.das@ibm.com}
}

\date{}
\renewcommand{\headeright}{}
\renewcommand{\undertitle}{}
\renewcommand{\shorttitle}{On observability and optimal gain design for distributed linear filtering and prediction}

\begin{document}

\maketitle

\begin{abstract}
This paper presents a new approach to distributed linear filtering and prediction. The problem under consideration consists of a random dynamical system observed by a multi-agent network of sensors where the network is sparse. Inspired by the \emph{consensus+innovations} type of distributed estimation approaches, this paper proposes a novel algorithm that fuses the concepts of consensus and innovations. The paper introduces a definition of distributed observability, required by the proposed algorithm, which is a weaker assumption than that of global observability and connected network assumptions combined together. Following first principles, the optimal gain matrices are designed such that the mean-squared error of estimation is minimized at each agent and the distributed version of the algebraic Riccati equation is derived for computing the gains. 
\end{abstract}


\section{Introduction}
The Kalman filter~\cite{kalman1960new, kalman1961new} is a celebrated solution for linear filtering and prediction of time-varying random fields observed with noisy measurements. The growing need for data privacy and robustness from centralized server failure, led to the development of distributed filtering and predictions algorithms. These new genre of distributed estimation algorithms~\cite{chong2017forty,ding2019survey} further reduces the large communication and computation overload at the centralized processors. Thereby, they position themselves as a critical framework for several applications namely multi-agent control~\cite{ghai2022regret}, indoor positioning \& navigation, state estimation in power grid, spatio-temporal environment or field monitoring~\cite{li2021cooperative}, connected vehicular network for traffic balancing \& accident aversion, collaborative target tracking~\cite{yan2018feedback} etc. 

In the literature, there are distributed estimators where the agents exchange information multiple number of times between each dynamics/observations iterates \cite{olfati2005distributed, olfati2007distributed, khan2008distributing, carli2008distributed, schizas2008consensus, ribeiro2010kalman, casbeer2009distributed}, so that average consensus occurs between observations. There are also single time-scale approaches \cite{das2015TSP, das2017TSP,khan2011coordinated,khan2013genericity,martins2012augmented,das2013ICASSP,das2013EUSIPCO} where the agents collaborate with their neighbors only once in between each dynamics/observation cycle. There are distributed Kalman filters~\cite{kar2011gossip, li2015distributed}  where the agents communicate among themselves using the Gossip protocol~\cite{dimakiskarmourarabbatscaglione-2010}, over a dynamic communication network~\cite{howard2021optimal} or over noisy/corrupted communication channels~\cite{he2021distributed,savas2017distributed,zhang2017distributed}. Some distributed estimators run a companion filter to estimate the global average of the pseudo-innovations~\cite{das2013Allerton,das2013Asilomar}, a modified version of the innovations.

There is a gap in the literature in providing for a distributed state estimation algorithm that is as general and robust as its centralized counterpart, the Kalman filter. In some papers, there are stricter assumptions on the local observation model or the communication network among the agents, for example neighborhood observability~\cite{olfati2009kalman, cattivelli2010diffusion} or an undirected connected graph~\cite{das2015TSP}. In practical distributed settings, these assumptions become a bottleneck. In other papers, there is an upper limit on the degree of instability on the system dynamics that a given observation-network model can handle with bounded MSE~\cite{khan2011networked}. Such algorithms become inapplicable to distributed process control domains where the underlying systems are inherently unstable and needs to be stabilized by appropriate control input. If the estimation algorithm fails to track the unstable system, then the design of a stabilizable control input become infeasible. The biggest gap is in the optimality of the distributed state estimates which leverages the maximum information available from the local observations and the estimates obtained from neighbors. The literature lacks in providing the optimal gain matrix design such that the algorithm yields yield minimum MSE estimates. 

This paper proposes a novel framework and an algorithm that addresses all these gaps in the distributed filtering and prediction literature. The algorithm is provided in Section~\ref{sec-algo},  where the consensus on the state estimates are treated as innovations and the gain matrices are designed appropriately. Such optimal gains, provided in Section~\ref{sec:gain}, yield field estimates with minimum mean-squared error (MMSE) at each agent under the assumption of distributed obervability at each agent. A new definition of distributed observability is introduced in Section~\ref{sec-obs} which is agnostic of the communication graph being directed or undirected and does not require the graph to be connected. First, we start with setting up the system-observation model framework in the following Section~\ref{sec-system}.

\section{System-Observation-Communication Model}
\label{sec-system}
The system under consideration follows a discrete-time, linear, and time-invariant state-space model
\begin{align}
\label{eqn:sys}
    \bx_{k} = F \bx_{k-1} + \bw_{k-1},
\end{align}
where, $\bx_{k} \in \mathbb{R}^{n}$ is the dynamic random state vector, $F \in \mathbb{R}^{n\times n}$ is the state transition matrix, $\bw_{k} \in \mathbb{R}^{n}$ is the system noise at all time $t=kT$ where T is the discrete-time step size and k is an integer time index. The system noise is white Gaussian noise with zero mean and covariance matrix $Q_k$, i.e., $\bw_k \sim \mathcal{N}(0, Q_k)$. The initial condition of the system,~$\bx_0$ is also Gaussian, that follows $\bx_0 \sim \mathcal N(\overline{\bx}_0, P^+_0)$.

The dynamic random state~\eqref{eqn:sys} is observed by a multi-agent network of $m$ agents (sensors). Each agent~$i$ observes only a few state variables and makes low dimensional measurements~$\bz_{i,k} \in \mathbb{R}^{p_i}$, such that $p_i << n, \forall i=1,\hdots,m$. The observations of the agents in the cyber layer is represented by a linear and time-invariant model
\begin{align}
\label{eqn:obs}
    \bz_{i,k} = H_{i} \bx_{k} + \bv_{i,k}, \qquad i=1,\hdots,m
\end{align}
where, $H_{i} \in \mathbb{R}^{p_i\times n}$ is the measurement matrix and~$\bv_{i,k} \in \mathbb{R}^{p_i}$ is the measurement noise. The measurement noise, at each agent~$i$, is also white Gaussian noise with zero mean and covariance matrix $R_{i,k}$, i.e., $\bv_{i,k} \sim \mathcal{N}(0, R_{i,k})$. The system noise, the measurement noise, and the initial condition $\{ \{\bw_k\}, \{\bv_{i,k}\}, \bx_0\}_{\forall i, k \geq 0}$  are uncorrelated random sequences.

The agents in the network layer exchange their measurements and current estimates with their neighbors. Formally, the agent communication network is represented by a simple (no self-loops nor multiple edges) and \emph{directed} graph $\mathcal{G = (V,E)}$, where $\mathcal{V} = \{i: i=1,\hdots,m\}$ is the set of agents and  $\mathcal{E} = \{(i,j): \exists \text{ an edge }j \rightarrow i\}$ is the set of local communication channels among the agents. We consider directed graph (one-way communications), that means our algorithm is easily extendable to undirected graphs (two-way communications) but the reverse is not always true. The adjacency matrix of $\mathcal{G}$ is denoted by $\ad = [a_{ij}] \in \mathcal{R}^{m\times m}$, where, 
\begin{align}
    \label{eqn:net}
    a_{ij} =  \left\{
\begin{array}{ll}
1,  & \mbox{if } \exists \text{ an edge }j \rightarrow i  \\
0, & \text{otherwise.}  \\
\end{array} \right. 
\end{align}
For details on graphs refer to \cite{chung1997spectral}. The communication network is sparse and time-invariant. For each agent~$i$, let's define the open and closed neighborhoods as:
\begin{align}
\label{eqn:open-nbhd}
    \Omega_i &= \{j|(i,j) \in  \mathcal{E}\} \\
    \label{eqn:closed-nbhd}
    \overline{\Omega}_i &=  \{i\} \cup \{j|(i,j) \in  \mathcal{E}\}.
\end{align}

As in the case of a classical optimal Kalman filter, each agent~$i$ in the framework knows the system model, $F$ and $Q_k$, the initial condition statistics, $\overline{\bx}_0$ and $P^+_0$, the parameters of its and neighbors' measurement models, $\{H_j, R_{j,k} : j \in \overline{\Omega}_i \}$, and the communication network model, $\mathcal{G}$ along with the adjacency matrix~$\ad$. Note that the time-invariant state-space is chosen for notational simplicity. All the derivations, assumptions and the results in this paper also hold for a time-varying state-space model $(F_k, H_{i,k}, R_{i,k}, Q_k)$.
\footnote{Further, we excluded the input to the system $\bu_k$, the control matrix $C_k$ and the system noise matrix~$\Phi_k$ as the analysis remains the same if we include them. A complete time-varying state-space equation would be, \\
\begin{align*}
  \bx_{k} &= F_{k-1} \bx_{k-1} + C_{k-1} \bu_{k-1} + \Phi_{k-1}\bw_{k-1} \\
  \bz_{i,k} &= H_{i,k} \bx_{k} + \bv_{i,k}, \qquad i=1,\hdots,m.
\end{align*}
}

\section{Distributed Observability}
\label{sec-obs}
Before we propose the distributed estimation algorithm and describe the design of the optimal gain, we introduce the notion of distributed observability, a measure of how well internal states of a system can be inferred from knowledge of its local measurements and interactions among agents in the network. Consider the physical system-observation-communication modeled by the {\bf \emph{state-space-network}} representaion~\eqref{eqn:sys}-\eqref{eqn:net}. The local observability matrix~$G_i \in \mathcal{R}^{np_i\times n}$ and the global observability matrix~$G\in \mathcal{R}^{\left( \displaystyle n \Sigma_{i=1}^{m}p_i \right) \times n}$ of the network are denoted by,
\begin{align}
    \label{eqn:obs-matrix}
    G_i = \begin{bmatrix} H_i \\ H_iF \\ H_iF^2 \\ \vdots \\ H_i F^{n-1} \end{bmatrix}, \; \forall~i\in\mathcal{V}; \qquad
    G =  \begin{bmatrix} G_1 \\ G_2 \\ \vdots \\ G_m \end{bmatrix}.
\end{align}

Let's define the {\bf \emph{connectivity matrix}}~$\tilde{\ad}$ of the network as,
\begin{align}
    \label{eqn:connectivity-matrix}
    \tilde{\ad} = I_m + \ad + \ad^2 + \hdots + \ad^{m-1}.
\end{align}
The element~$[\ad^q]_{i,j}$ of the matrix,~$\ad^q \; \forall q \in \mathcal{Z}^+$, gives the number of directed walks of length~$q$ from agent~$j$ to agent~$i$. Then, the connectivity matrix is a non-negative matrix,~$\tilde{\ad} \geq 0$, and its elements~$[\tilde{\ad}]_{i,j}=\tilde{a}_{i,j}$ denote the total number of walks (of any length $<m$) from node $j$ to node $i$. 
\begin{remark}
\label{rem:connectivity-matrix}
If there exists $i,j$ such that $\tilde{a}_{i,j}=0$, then there doesn't exist any path from $j$ to $i$ and would imply that the graph is not connected. The agent communication network, i.e., the directed graph, is connected\footnote{For a \emph{fully} connected network, $I_m + \ad > 0$.} if the connectivity matrix, defined in~\eqref{eqn:connectivity-matrix}, is a positive matrix, i.e., $\tilde{\ad}>0$.
\end{remark}
\begin{remark}
\label{rem:advantage}
Although most literature in distributed estimation requires a connected graph, but connected graph is not a necessary condition for the distributed observability (defined below) and is also not required for the distributed estimation algorithm proposed in this paper.
\end{remark}
For the defition of distributed observability, let the quantity~$\tilde{\ad}_i$ denote the~$i^{\text{th}}$ row of the matrix~$\tilde{\ad}$ and the symbol~$\bullet$ denote the face-splitting product of matrices (transposed Khatri–Rao product).
\begin{definition}[Distributed Observability]
If the row rank of the distributed observability matrix~$\mathcal{O}_i$, defined as,
\begin{align}
    \label{eqn:dist-obs}
    \mathcal{O}_i = \tilde{\ad}_i \bullet G = 
    \begin{bmatrix} \tilde{a}_{i,1} \\ \tilde{a}_{i,2} \\ \vdots \\ \tilde{a}_{i,m} \end{bmatrix} \bullet
    \begin{bmatrix} G_1 \\ G_2 \\ \vdots \\ G_m \end{bmatrix}
    = \begin{bmatrix} \tilde{a}_{i,1}\otimes G_1 \\ \tilde{a}_{i,2}\otimes G_2 \\ \vdots \\ \tilde{a}_{i,m}\otimes G_m \end{bmatrix}
\end{align}
is equal to $n$, then the system is distributedly observable at agent~$i$.
\end{definition}
Note that,~$\mathcal{O}_i \in \mathcal{R}^{\left( \displaystyle n \Sigma_{i=1}^{m}p_i \right) \times n}$. The requirement of invertibility (or full-rank) of the distributed observability Gramian, $\mathcal{O}_i^T \mathcal{O}_i$, is an equivalent alternative of the distributed observability definition, i.e., rank$( \mathcal{O}_i^T \mathcal{O}_i ) = n$. 
\begin{assumption}
\label{assm:dist-obs}
The state-space-network model ~\eqref{eqn:sys}-\eqref{eqn:net} is distributedly observable at all agents in the network.
\end{assumption}
This is a crucial assumption in this paper that ensures that the proposed distributed estimator converges with bounded mean-squared error\footnote{A slightly weaker criteria defined as distributed detectability suffices for the convergence of the proposed algorithm. It requires only the unstable states to be observable.}. Note that there is no requirement for the system to be stable or the network to be connected. 
%
%
With the distributed observability definition and the Assumption~\ref{assm:dist-obs}, we state the algorithm in the next section followed by the optimal gain design.

\section{Distributed Estimation Algorithm}
\label{sec-algo}
At time~$k$ and agent~$i$, let us denote the filter and prediction estimates of the system by~$\xf_{i,k}$ and~$\xp_{i,k}$, respectively, and the filter and prediction error covariance matrices by~$\pf_{i,k}$ and~$\pp_{i,k}$, respectively. The prediction and filtering updates of the distributed estimation algorithm at agent~$i$ are:
\begin{align}
    \label{eqn:xp-update}
    \xp_{i,k} &= F \xf_{i,k-1} \\
    \label{eqn:pp-update}
    \pp_{i,k} &= F\pf_{i,k-1} F^T + Q_k \\ 
    \label{eqn:gain-update}
    \bk_{i,k} \! \! &= \Sigma_{x, y_i} \Sigma^{-1}_{y_i} \\
    \label{eqn:xf-update}
   \xf_{i,k} &\! = \! \xp_{i,k} \!+\!\!\! \sum_{j \in \Omega_i}\!\! B_{ij,k} \!\! \left( \xp_{j,k} \!-\! \xp_{i,k}\right) + \sum_{j \in \overline{\Omega}_i} \!\! M_{ij,k} \!\! \left( \bz_{j,k} \!-\! H_j \xp_{i,k} \right) \\
    \label{eqn:pf-update}
    \pf_{i,k} &= \pp_{i,k} - \bk_{i,k} \Sigma^T_{x, y_i}
\end{align}

%
%
where,~$\bk_{i,k} \in \mathbb{R}^{ n \times \left( \sum_{j \in \cn_i}p_j + n|\on_i| \right)}$ is the distributed gain matrix. The covariance matrices~$\Sigma_{x, y_i}$ and~$\Sigma_{y_i}$ are derived in the optimal gain design Section~\ref{sec:gain}. The local consensus weight matrices,~$B_{ij,k} \in \mathbb{R}^{n \times n}$, and the local innovation weight matrices,~$M_{ij,k} \in \mathbb{R}^{n \times p_j}$, are obtained from the distributed gain matrix~$\bk_{i,k}$ as shown in~\eqref{eqn:gain2weights}.

The equations~\eqref{eqn:xp-update}-\eqref{eqn:pf-update} represents the proposed distributed estimation algorithm, where the minimized MSE prediction and filter estimates $\xp_{i,k}$ and $\xf_{i,k}$ are the conditional means,
\begin{align}
\label{eqn:MMSE_xp}
\xp_{i,k} &= \mathbb{E} \left[ \bx_k \; | \; \{ \bz_{j,k-1} \}_{j \in \cn_i}, \{ \xf_{j,k-1} \}_{j \in \on_i}\right] \\
\label{eqn:MMSE_xf}
\xf_{i,k} &= \mathbb{E} \left[ \bx_k \; | \; \{ \bz_{j,k} \}_{j \in \cn_i}, \{ \xp_{j,k} \}_{j \in \on_i}\right].
\end{align}

The filter update~\eqref{eqn:xf-update} is visually similar to \emph{consensus+innovations} type of algorithms\footnote{The distributed estimator is inspired by the pseudo-innovations, pseudo-observations and pseudo-state approaches that are summarized in~\cite{das2016distributed, das2017distributed}.}. But functionally this algorithm fuses the concepts of consensus and innovations by treating the consensus on the state estimates as innovations along with the local innovations of the agent and its neighbors. To best represent this functionality, the filter update~\eqref{eqn:xf-update} is re-written with the local innovation term,~$\by_{i,k}$ at agent~$i$, as:
\begin{align}
    \label{eqn:xf-inno}
    \xf_{i,k} =& \xp_{i,k} + \bk_{i,k} \underbrace{\begin{bmatrix} 
    \bz_{j_1,k} - H_{j_1} \xp_{i,k} \\ \cdot\cdot \\ \bz_{i,k} - H_i \xp_{i,k} \\ \cdot \cdot \\ \bz_{j_{|\cn_i|},k} - H_{j_{|\cn_i|}} \xp_{i,k} \\
    \xp_{j_1,k} - \xp_{i,k} \\ \vdots \vdots \\ \xp_{j_{|\on_i|},k} - \xp_{i,k}
    \end{bmatrix}}_{\by_{i,k}},  \qquad \text{where,}\\
    \label{eqn:gain2weights}
    \bk_{i,k} =& \begin{bmatrix} M_{ij_1,k}, \cdot \cdot, M_{ii,k}, \cdot \cdot, M_{ij_{|\cn_i|},k}, B_{ij_1,k}, \cdots, B_{ij_{|\on_i|},k} \end{bmatrix}, 
\end{align}
$\{j_1, \cdot \cdot, i, \cdot \cdot, j_{|\cn_i|} \} = \cn_i$ and  $\{j_1, \cdots, j_{|\on_i|}\} = \on_i$.

The innovation sequences~$\{\by_{i,k}\}_{\forall i, k\geq0}$ are Gaussian random vectors, uncorrelated and are with zero mean, $\mathbb{E}[\by_{i,k}]=0$, $\forall i, k\geq 0$. These innovation terms are key to the optimal design of the gain matrices. 

\section{Error Analysis}
\label{sec:error}
Now we derive the predictor and filter error terms,~$\ep_{i,k} \in \mathbb{R}^n$ and~$\ef_{i,k} \in \mathbb{R}^n$, respectively, at each agent~i,
\begin{align}
    \label{eqn:ep}
    \ep_{i,k} &= \bx_k - \xp_{i,k} \\
    \label{eqn:ef}
    \ef_{i,k} &= \bx_k - \xf_{i,k}.
\end{align}
The error processes~$\ep_{i,k}$ and~$\ef_{i,k}$ are unbiased, i.e, they are zero mean at all agents and for all time indices, $\mathbb{E}[\ep_{i,k}]=0$ and $\mathbb{E}[\ef_{i,k}]=0$, $\forall i, k\geq 0$. The error processes follows: $\ep_{i,k} \sim \mathcal{N} ({\bm 0}_n, \pp_{i,k})$ and $\ef_{i,k} \sim \mathcal{N} ({\bm 0}_n, \pf_{i,k})$. This shows that the distributed prediction~$\xp_{i,k}$ and filtering~$\xf_{i,k}$ estimates provided by this algorithm are unbiased.

From equation~\eqref{eqn:xf-inno} and using~\eqref{eqn:obs}, the innovations are expanded as:
\begin{align*}
   \by_{i,k} =  \begin{bmatrix} 
    H_{j_1} \bx_{k} + \bv_{j_1,k} - H_{j_1} \xp_{i,k} \\ \cdot\cdot \\  H_{i} \bx_{k} + \bv_{i,k} - H_i \xp_{i,k} \\ \cdot \cdot \\  H_{j_{|\cn_i|}} \bx_{k} \!+\! \bv_{j_{|\cn_i|},k} \!-\! H_{j_{|\cn_i|}} \xp_{i,k} \\
    \xp_{j_1,k} - \bx_k + \bx_k - \xp_{i,k} \\ \vdots \vdots \\ \xp_{j_{|\on_i|},k} - \bx_k + \bx_k - \xp_{i,k}
    \end{bmatrix}
    =  \underbrace{ \begin{bmatrix} 
    H_{j_1} \\ \cdot\cdot \\  H_{i} \\ \cdot \cdot \\  H_{j_{|\cn_i|}} \\ {\bm 0}_{n\times n} \\ \vdots \vdots \\ {\bm 0}_{n\times n} \end{bmatrix} }_{\tilde{H}_i} \! \ep_{i,k} \!\!+\!\! 
    \underbrace{ \begin{bmatrix} 
    \bv_{j_1,k}\\ \cdot\cdot \\  \bv_{i,k} \\ \cdot \cdot \\  \bv_{j_{|\cn_i|},k} \\
    \ep_{i,k} - \ep_{j_1,k} \\ \vdots \vdots \\ \ep_{i,k} \!\! - \!\! \ep_{j_{|\on_i|},k}
    \end{bmatrix} }_{\delta_{i,k}}
\end{align*}
where,~$\tilde{H}_i \in \mathbb{R}^{ \left( \sum_{j \in \cn_i}p_j + n|\on_i| \right) \times n}$ are the local innovation matrices and the~$\delta_{i,k} \in \mathbb{R}^{\sum_{j \in \cn_i}p_j + n|\on_i|}$ are the local innovation noises at each agent~$i$. In compact notation, the dynamics of the local innovations are represented by,
\begin{align}
    \label{eqn:inno-dynamics}
    \by_{i,k} = \tilde{H}_i \ep_{i,k} + \delta_{i,k}, \qquad \forall i, k\geq 0.
\end{align}
The local innovation noises~$\delta_{i,k}$ are Gaussian random vectors with zero mean and let the variance be denoted by~$\Delta_{i,k}$, i.e., $\delta_{i,k} \sim \mathcal{N}({\bm 0}, \Delta_{i,k})$. 
Using the equations~\eqref{eqn:sys}, \eqref{eqn:xp-update}, \eqref{eqn:xf-inno} and \eqref{eqn:inno-dynamics} on the predictor and filter errors~\eqref{eqn:ep} and~\eqref{eqn:ef}, their dynamics take the form:
\begin{align}
    \ep_{i,k} &= F \bx_{k-1} + \bw_{k-1} - F \xf_{i,k-1} \nonumber \\
    \label{eqn:ep-dyn}
    & = F \ef_{i,k-1} + \bw_{k-1} \\
    \ef_{i,k} &= \bx_k - \xp_{i,k} - \bk_{i,k} \by_{i,k} \nonumber \\ 
    &= \ep_{i,k} - \bk_{i,k}\tilde{H}_i \ep_{i,k} - \bk_{i,k}\delta_{i,k} \nonumber \\
    &= (I_n - \bk_{i,k}\tilde{H}_i)\ep_{i,k} - \bk_{i,k}\delta_{i,k} \nonumber\\
    \label{eqn:ef-dyn}
    \ef_{i,k} &= (F \!-\! \bk_{i,k}\tilde{H}_i F)\ef_{i,k-1} \!+\! (I_n \!-\! \bk_{i,k}\tilde{H}_i)\bw_{k-1} \!-\! \bk_{i,k}\delta_{i,k}
\end{align}
Given that the predictor and filter errors are zero-mean, the recursive updates of the evolution of their covariances,~\eqref{eqn:pp-update} and~\eqref{eqn:pf-update} are derived below. 

\begin{align}
    \pp_{i,k} &= \mathbb{E} [\ep_{i,k} {\ep}^{T}_{\!\!i,k}] \!=\! \mathbb{E}[ (F \ef_{i,k-1} \!\!+\!  \bw_{k-1}) (F \ef_{i,k-1} \!\!+ \!  \bw_{k-1})^T ] \nonumber \\
    \label{eqn:pp-update-deriv}
    &= F \pf_{i,k-1} F^T + Q_k \\
    \pf_{i,k} &= \mathbb{E} [\ef_{i,k} {\ef}^{T}_{i,k} ] = \mathbb{E} [ (\ep_{i,k} - \bk_{i,k} \by_{i,k}) (\ep_{i,k} - \bk_{i,k} \by_{i,k})^T ] \nonumber \\
    &= \pp_{i,k} - \bk_{i,k}\Sigma_{x, y_i}^T - \Sigma_{x, y_i}\bk_{i,k}^T + \bk_{i,k}\Sigma_{y_i}\bk_{i,k}^T \nonumber \\
    \label{eqn:pf-update-deriv}
    &= \pp_{i,k} - \bk_{i,k}\Sigma_{x, y_i}^T
\end{align}
Note that the expectations of the cross-terms in~\eqref{eqn:pp-update-deriv} are zero, which can be shown by using techniques similar to the ones presented in~\cite{das2017TSP}. The term~$\mathbb{E}[ \ep_{i,k}\by_{i,k}^T] = \Sigma_{x, y_i}$ as shown later in~\eqref{eqn:equi-cross-cov}. Finally,~$\eqref{eqn:pf-update-deriv}$ was obtained by substituting~$\bk_{i,k}$ with~$\Sigma_{x, y_i}\Sigma_{y_i}^{-1}$.

The convergence properties of the distributed estimator~\eqref{eqn:xp-update}-\eqref{eqn:pf-update} is determined by the dynamics of the filter and prediction error processes,~\eqref{eqn:ef-dyn} and~\eqref{eqn:ep-dyn}. If the error dynamics are asymptotically stable, then the error processes have asymptotically bounded error covariances that in turn guarantee the convergence of the distributed algorithm. Note that if the dynamics of the filter error processes,~$\ef_{i,k} \; \forall \; i$, are asymptotically stable, then the dynamics of the prediction error processes,~$\ef_{i,k} \; \forall \; i$ are also asymptotically stable. That is why the dynamics of only one of the error processes is typically studied and in this paper we consider the filter error processes.

For the distributed estimator to converge with bounded mean-squared error (MSE), the filter error~\eqref{eqn:ef-dyn} needs to be asymptotically stable, i.e., the spectral radius of the error's dynamics matrix has to be less than one,~$\rho(F - \bk_{i,k}\tilde{H}_i F) < 1$. Given that the \emph{state-space-network} model~\eqref{eqn:sys}-\eqref{eqn:net} satisfies the Distributed Observability criteria~\eqref{eqn:dist-obs}, it guarantees that there exists gain matrices~$\bk_{i,k}$ at each agent~$i$ such that ~$\rho(F - \bk_{i,k}\tilde{H}_i F) < 1$. This leads to the next section where a design of the optimal gain matrices are provided.

\section{Optimal Gain Design}
\label{sec:gain}
The asymptotic stability of the error dynamics guarantees convergence of distributed estimation algorithm~\eqref{eqn:xp-update}-\eqref{eqn:pf-update} and bounded MSE, but we aim to to design the gain matrices~$\bk_{i,k}$ such that the MSE is not only bounded but also minimum.

Since the zero-mean innovation sequences~$\{\by_{i,k}\}_{\forall i, k\geq0}$ are Gaussian and uncorrelated, they are independent random vectors. By applying Gauss-Markov theorem on equation~\eqref{eqn:gain2weights}, the gain matrices that minimizes the MSE of the filter and prediction estimates are given by,
\begin{align}
    \bk_{i,k} &= \Sigma_{x, y_i} \Sigma^{-1}_{y_i}, \qquad \text{where,} \nonumber \\
    \Sigma_{x, y_i} &= \mathbb{E}[ (\bx_k - \overline{\bx}_k)\by_{i,k}^T] = \mathbb{E}[ (\bx_k - \xp_{i,k} + \xp_{i,k} - \overline{\bx}_k)\by_{i,k}^T] \nonumber \\
    \label{eqn:equi-cross-cov}
    &= \mathbb{E}[ \ep_{i,k}\by_{i,k}^T] =  \mathbb{E}[ \ep_{i,k}( \tilde{H}_i \ep_{i,k} + \delta_{i,k} )^T] \\
    \label{eqn:cross-inno-x}
    &= \pp_{i,k}\tilde{H}_i^T + \Sigma_{\epsilon_i, \delta_i}\\
    \text{and,}& \nonumber \\ 
    \Sigma_{y_i} &= \mathbb{E} [\by_{i,k} \by_{i,k}^T] = \mathbb{E} [ (\tilde{H}_i \ep_{i,k} + \delta_{i,k}) (\tilde{H}_i \ep_{i,k} + \delta_{i,k})^T ] \nonumber \\
    &= \tilde{H}_i \pp_{i,k}\tilde{H}_i^T + \Delta_{i,k} + \tilde{H}_i \Sigma_{\epsilon_i, \delta_i} + \Sigma_{\epsilon_i, \delta_i}^T \tilde{H}_i^T.
\end{align}

The fact that~$\mathbb{E}[ (\xp_{i,k} - \overline{\bx}_k)\by_{i,k}^T] =0$ was utilized to obtain~\eqref{eqn:equi-cross-cov}. Now we employ the relations that~$\mathbb{E}[\ep_{i,k}\bv_{j,k}^T] =0 \; \forall\; j \in \cn_i$, to derive the two covariance quantities~$\Sigma_{\epsilon_i, \delta_i}$ and~$\Delta_{i,k}$.
\begin{align}
    \Sigma_{\epsilon_i, \delta_i} &= \mathbb{E}[ \ep_{i,k}\delta_{i,k}^T] \nonumber \\
    \label{eqn:sig-eps-delta}
    & = \begin{bmatrix}
     {\bm 0}_{n,p_{j_1}} \!\!\!\!\!&\!\! \!\!\cdot\cdot \!\!\!&\!\!  {\bm 0}_{n,p_{i}} \!\!\!&\!\!\!\! \cdot \cdot \!\!\!&\!\!\!  {\bm 0}_{n,p_{j_{|\cn_i|}}} \!\! \vdots (\pp_{i,k} \!\!-\!\! \pp_{ij_1,k})  \cdots (\pp_{i,k} \!\!-\!\! \pp_{ij_{|\on_i|},k})
    \end{bmatrix} \\
    \Delta_{i,k} &= \mathbb{E}[ \delta_{i,k} \delta_{i,k}^T] \nonumber \\
    \label{eqn:delta-cov}
    & = \text{blkdiag} \begin{Bmatrix} \text{blkdiag}\{ R_{j,k} \}_{j \in \cn_i},  \left[ [ \pp_{i,k} \!-\! \pp_{ji,k} \!-\! \pp_{il,k} \!+\! \pp_{jl,k}]_{j,l \in \on_i} \right] \end{Bmatrix}
\end{align}
where, blkdiag means a block-diagonal matrix. With the two expressions~$\Sigma_{\epsilon_i, \delta_i}$ and~$\Delta_{i,k}$ in~\eqref{eqn:sig-eps-delta}-\eqref{eqn:delta-cov}, the optimal gain matrices for the distributed estimator at each agent turns into:
%
\begin{align}
    \label{eqn:gain-full}
    \bk_{i,k} = \left( \pp_{i,k}\tilde{H}_i^T + \Sigma_{\epsilon_i, \delta_i} \right) \Big( \tilde{H}_i \pp_{i,k}\tilde{H}_i^T + \Delta_{i,k} + \tilde{H}_i \Sigma_{\epsilon_i, \delta_i} + \Sigma_{\epsilon_i, \delta_i}^T \tilde{H}_i^T \Big)^{-1}
\end{align}
%
Note that the gain matrices will be very sparse at each agent. To alleviate challenges in tracking of the complete network error covariances,~\cite{sebastian2021all} presents a certifiable optimal distributed filter that performs optimal fusion of estimates under unknown correlations by a particular tight Semidefinite Programming
(SDP) relaxation. Further, given that the matrices does not depend on the measurements, they all can be pre-computed and stored at each agent. 

Combining equations~\eqref{eqn:pp-update-deriv} and~\eqref{eqn:pf-update-deriv}, we get
\begin{align}
\label{eqn:dare}
    \pf_{i,k} = F \pf_{i,k-1} F^T + Q_k - \bk_{i,k}\Sigma_{x, y_i}^T
\end{align}
Equation~\eqref{eqn:dare}, once substituted with~\eqref{eqn:cross-inno-x}-\eqref{eqn:gain-full} to express in terms of system parameters, will yield a recursive iteration of the filter error covariance matrix which is the distributed version of the discrete algebraic Riccati equation for the proposed distributed estimation algorithm. The initial condition of the covariances are~$\pf_{i,0} = \pf_{0} \; \forall \; i$ and~$\pf_{ij,0} = \pf_{0} \; \forall \; i, j\in \on_i$. 

Under the distributed observability Assumption~\ref{assm:dist-obs}, the Riccati equation has an asymptotic solution at each agent which is positive definite when started with a symmetric positive semi–definite matrix. This solution, which we will designate by~$\pf_{i,\infty}$ is the fixed point of equation~\eqref{eqn:dare}. For the linear time invariant problems (assuming distributed observability), the steady state filter is asymptotically stable~$\rho(F - \bk_{i,\infty}\tilde{H}_i F) < 1$, i.e., the closed loop filter matrix~$F - \bk_{i,\infty}\tilde{H}_i F$ has all poles inside the unit circle, regardless of F being or not asymptotically stable. To save on the storage burden, the steady state gain matrix~$\bk_{i,\infty}$ could be used at each agent~$i$ for all the iterations. This may not yield distributed estimates with minimum MSE, but will certainly provide estimates with bounded MSE.


\section{Conclusions}
The primary contributions of this paper are: (i) providing a novel and meaningful definition for distributed observability; (ii) introducing a new class of distributed state estimation algorithm that treats consensus on neighbors' estimates as innovations; and (iii) designing the gain matrices for the distributed estimator such that the algorithm is optimal, i.e., it yields minimum MSE estimates at all agents. The algorithm, derivations and error analyses presented in this paper resolves most of the challenges related to convergence and optimality of distributed state estimation. The methodologies proposed in this paper have the potential to serve as the backbone for several downstream research challenges including sensor placement in a multi-agent network, adaptation to node or communication failures and other related research problems.  

\bibliographystyle{unsrt}
\bibliography{references}

\end{document}